\documentclass{aa}   

\usepackage{graphicx}
\usepackage{psfig, epsfig}

\begin{document}
\title {Is evidence for enhanced mass transfer during dwarf-nova 
outbursts well substantiated?}

\author{Yoji Osaki \inst{1}
\and Friedrich Meyer \inst{2}}
\offprints{Yoji Osaki; osaki@net.nagasaki-u.ac.jp}

\institute{Faculty of Education, Nagasaki University, Nagasaki
 852-8521, Japan 
\and
 Max-Planck-Institut f\"ur Astrophysik, Karl Schwarzschild Str. 1, 
D-85740 Garching, Germany}

\date{Received: / Accepted:}

\abstract{
Outburst mechanisms of SU UMa-type dwarf novae are discussed. Two competing 
models were proposed; a pure disk instability model called the 
thermal-tidal instability model (TTI model) and the enhanced mass transfer 
model (EMT model). Observational evidence for enhanced mass transfer 
from the secondary star during outbursts is critically examined. It is 
demonstrated that most evidence for enhanced mass transfer is not well 
substantiated. 
Patterson et al. (2002) have recently claimed to have found 
evidence for enhanced mass transfer during the 2001 outburst of WZ Sge.  
We show that their evidence  is probably due to a misinterpretation of
their observed light
curves. Our theoretical analysis also shows that irradiation during
outburst should not affect the mass transfer rate. 
A refinement of the TTI model is proposed that can explain
why superhumps appear a few days {\it after} the superoutburst maximum 
in some SU UMa stars. 
We present our own interpretation of the overall development of 
the 2001 outburst of WZ Sge based on the TTI model that does not
require the assumption of an unproved enhanced mass transfer. 
\keywords{ accretion, accretion disks -- binaries: close -- 
novae, cataclysmic variables -- stars: dwarf novae --
stars: individual: WZ Sge}
}
\titlerunning {Enhanced mass transfer during dwarf-nova outbursts ?}
\maketitle

\section{Introduction}

SU UMa stars are dwarf novae of short orbital periods that show two distinct 
types of outbursts, called short normal outburst and long superoutburst 
(see, e.g., Warner (1995) for dwarf novae in general and SU UMa stars in 
particular). 
In ordinary SU UMa stars several short normal outbursts are sandwiched 
between two consecutive superoutbursts. The cycle from one superoutburst
to the next is called supercycle. 
During the superoutburst periodic humps called superhumps 
appear that repeat with a period slightly longer than the binary
orbital period. 
The superhump phenomenon in SU UMa stars is now well understood in terms of 
the tidal instability  (Whitehurst 1988; Hirose \& Osaki 1990; Lubow 1991): 
superhumps are produced by periodic tidal stressing of the eccentric
precessing accretion disk which is formed by the 3:1 resonance tidal
instability.  

The short normal outburst of SU UMa stars is thought to be essentially 
the same as the outburst of U Gem stars and is believed to be caused by  
the thermal instability in the accretion disk. For the superoutburst 
and supercycle of SU UMa stars Osaki (1989) has proposed the thermal-tidal 
instability model (TTI model) in which the ordinary thermal 
instability is coupled with the tidal instability. 
Based on the TTI model, Osaki (1996) has presented a unification model 
for outbursts of dwarf novae in which a rich variety of outburst behaviors of 
non-magnetic cataclysmic variable stars is understood within the general 
framework of the disk instability model. 
 
Although the TTI model is basically successful in explaining 
the superoutburst phenomenon of SU UMa stars, there still remains 
some deficiency. Hellier (2001) has suggested 
a minor modification to the TTI model in the case of binary 
systems with extremely low mass ratio. 
Moreover, objections to the TTI model were raised by 
Smak (1996, 2000) and by Lasota's group (see Lasota (2001) for 
a review of his group's view of this problem). 
They argue that enhanced mass transfer, likely caused 
by irradiation of the secondary star by the central source of radiation, 
is responsible for the superoutburst of SU UMa stars (EMT model). 
Historically speaking, an enhanced mass transfer model by an irradiated 
secondary  was proposed by Osaki (1985) to explain superoutbursts of
SU UMa stars but this model was later abandoned 
by its author in favor of the TTI model (Osaki 1996).  
On page 486  of his (2001) review, Lasota made the following statement:
``the main deficiency of the TTI model is that it neglects a property of SU UMa 
stars and dwarf novae outbursts in general: the {\it observed} enhancement 
of the mass-transfer rate during outbursts. "

The purpose of this paper is to examine critically whether
Lasota's statement, i.e., the enhancement of the mass-transfer 
during outbursts, is  observationally and theoretically substantiated. 
Patterson et al. (2002) recently presented extensive photometry 
observations of the 2001 superoutburst of WZ Sge.  
 From eclipse observations during the main outburst 
of WZ Sge in 2001 they claim that the mass transfer rate was enhanced 
by a factor of up to 60 from its quiescent value. 
We also discuss these observations and their interpretation. 

In Sect. 2, we examine the various claims for enhanced mass 
transfer and we show that most evidence is not well founded.
Our theoretical analysis in Sect. 3 shows severe problems for
any model that claims an effect of irradiation on the mass transfer
rate in outburst. 
In response to Smak's (1996, 2001) criticism of the TTI model, we propose 
in Sect. 4 a refinement of the TTI model that explains why 
superhumps appear a few days {\it after} the superoutburst maximum 
in some SU UMa stars. 
After thus having demonstrated that enhanced 
mass transfer during outburst is not well substantiated, we present 
our own interpretation for the overall development of the 2001 
outburst of WZ Sge in Sect. 5. In the Appendix, we critically discuss 
the new interpretation of the early hump phenomenon by Kato (2002). 
A conclusion of the paper is given in Sect. 6.

\section {Examination of claimed observational evidence for enhanced mass 
transfer during outburst}

\subsection {Early hump phenomenon in WZ Sge}
 
Periodic humps appeared during the first 10 days of the 1978 outburst of WZ Sge 
that repeated with the orbital period. These humps have been called either 
``outburst orbital humps" or ``early superhumps". Here we call them 
simply "early humps" (see the discussion in Osaki and Meyer, 2002).   
It is interesting to note that the early hump phenomenon was used 
as evidence for enhanced mass transfer in two different ways 
by Patterson et al. (1981) and by Smak (1993). 

Patterson et al. (1981) interpreted the humps repeating with the orbital 
period in the first 10 days of the 1978 outburst as brightening 
of the hot spot caused by enhanced mass transfer. 
However they had a difficulty with this interpretation 
since the position of the hot spot would have had to move azimuthally 
by 60 degrees from its quiescent position. The double hump nature of 
early humps observed in the 2001 outburst of WZ Sge also is difficult 
to explain by a hot spot brightening (Osaki \& Meyer 2002).   

 Smak (1993) on the other hand took the early humps observed 
in the 1978 WZ Sge outburst as evidence for a heated secondary. But also 
he had a difficulty with the phase of maximum visibility of the
irradiated surface of 
the secondary star. Nevertheless, he calculated the temperature of the
heated surface of the secondary star by assuming that radiation 
of the central source intercepted by the secondary star is 
thermalized in the photosphere and raises the surface temperature 
of the secondary star. This point will be discussed in the next section. 

However, if the early hump phenomenon is of disk origin 
due to a non-axisymmetric structure produced by tidal effects 
in our 2:1 resonance model (Osaki \& Meyer 2002) or 
in Kato 's model (Kato 2002),  the evidence for a brightened hot spot 
 or for a heated  secondary star  
disappears. Thus the early hump phenomenon of WZ Sge stars can not
be taken as evidence for enhanced mass transfer from the secondary star. 

\subsection {Enhanced humps in normal outbursts presented by Vogt 
(1983)}
   
One piece of evidence for enhanced mass transfer referred
to by Smak (1996) was 
found in the observational review of VW Hyi by Vogt
(1983). Vogt noted that significant enhancements in the hump amplitude 
during short eruptions were present in only four of nine cases and 
that all four cases were confined to those occurring less than 40 days 
from the next superoutbursts, summarized in Fig. 4 of his paper. 

In Vogt's picture, enhanced humps observed in maximum or decline 
 of normal short eruptions of VW Hyi were due to brightening
of the hot 
spot and the enhanced amplitudes of the hot spot were evidence 
for an enhanced mass transfer from the secondary star. We question 
Vogt's interpretation of the enhanced orbital humps in these 
four cases. Instead we suggest that the observed humps 
rather are of superhump nature in an embryonic form. 
By superhumps in an embryonic form we mean superhumps in the 
early stage of their growth. Such embryonic 
superhumps  were observed in precursor outbursts in T Leo by Kato (1997) and 
in V436 Cen by Semeniuk (1980).

Observations discussed below were done a quarter of a century ago and 
this question can be easily settled by new observations with CCD. We 
urge observers to make such observations. 

In fact, the normal outburst, claimed by Vogt (1983) to have 
exhibited a 15 times increase in hot spot brightness, 
was the precursor outburst of the May 1977 superoutburst of VW Hyi observed
by Marino \& Walker (1979). If we compare the light curve of this 
superoutburst (Fig. 4 of Marino \& Walker 1979) with those of the 1993 
January superoutburst of T Leo observed by T. Kato (Figs. 1 and 2 
of Kato 1997), we find they are very similar, exhibiting basically 
the same phenomenon. Kato (1997) has found that the hump period during 
decline from the precursor outburst of T Leo in January 
1993 is  2.4\% longer than the orbital period and thus these humps are 
embryonic superhumps rather than orbital humps. In the case of the 
precursor outburst in the decline stage of the May 12, 1977 outburst 
observed by Marino 
and Walker, only three hump maxima were seen and it would be difficult 
to determine their exact period in order to decide whether they are enhanced 
orbital humps or embryonic superhumps.

Marino and Walker (1979) found another hump-like feature in the decline 
of the 1978 October normal outburst of VW Hyi which occurred 
160 days after the previous superoutburst (the mean supercycle of VW Hyi 
is 180 days), but they were puzzled because the observed hump feature 
looked like a superhump with a period similar to that of superhumps 
rather than that of orbital humps. 
We interpret these humps as indeed embryonic superhumps 
that failed to develop into fully-grown superhumps and superoutburst. 
Two of the four cases claimed to exhibit an enhanced hot spot 
by Vogt (1983) were the same two observations by Marino and Walker 
(1979) discussed above.

The other two cases were those observed by Haefner et al. (1979). 
These two are outbursts designated as No. 280 and No. 281 which occurred 
around Nov. 15 and Nov. 26, 1974 while the next superoutburst 
designated as No. 282 occurred around Dec 7/8 in 1974. With respect to
the outburst 
No. 281, light curves for three consecutive nights were shown in Fig. 4 
of Haefner et al. (1979). The phases of the hump maxima in the first 
and third night agreed with those expected from the quiescent hump 
ephemeris while the phase of the biggest hump in the second night 
was shifted by 0.5 from that expected. Haefner et al. (1979) 
interpreted this hump maximum observed in the second night as 
a secondary maximum sometimes seen in quiescent orbital humps. 
However, another interpretation is possible. If we combine 
the hump maxima in all three nights in one ephemeris, we would get 
a hump period exceeding the orbital period by 3.8\% (i.e., a beat period 
of 2 days between the orbital and superhump periods). Thus these humps 
in outburst No. 281 of VW Hyi could well be interpreted as embryonic 
superhumps which failed to develop into fully-grown superhumps 
rather than as orbital humps. No light curves for the outburst 
No. 280 of VW Hyi are available in published form. However,  
the enhancement of hump amplitude in this outburst claimed by Vogt 
(1983) was rather small (the smallest among the four cases, 
i.e., enhancement less than 5 in Vogt 1983) and we do not  
discuss this outburst any further. 

In conclusion, the evidence for enhanced mass transfer presented 
by Vogt (1983) is not sufficient to substantiate the claim 
that the mass transfer rate is enhanced in the maximum and decline stages 
of normal outbursts if they occur near the next superoutburst. 

\subsection {Eclipse observations of the 2001 superoutburst of WZ 
Sagittae and Patterson's et al. (2002) interpretation for enhanced 
mass transfer}

Patterson et al. (2002) have presented extensive photometric
observations of the 2001 outburst of WZ Sge. During a first stage of
about 12 days early humps (outburst orbital humps or ``OOH'' in
Patterson et al. 2001) were visible. This stage is discussed in Osaki \& Meyer
(2002). In the following stage common superhumps were observed. Eclipses
suddenly appeared with the growth of the superhumps and were visible
all the way from the main outburst stage through the so called echo
outburst stage up to the final decline.

By synchronous summation of the superhump light curves at the orbital
period through out the beat phase between orbital and superhump period
Patterson et al. (2002) obtained ``orbital lightcurves'' that looked
like those of the quiescent orbital hump (lowest right panel of their
Fig. 7 and left panel of their Fig. 16). Therefore the authors interpreted
them as hot spot light curves and the eclipses observed as eclipses of
the mass transfer hot spot. By comparing intensity amplitudes of
the hump light they concluded that the mass transfer rate during this
stage had increased by a factor up to 60 above the quiescent
level. This would strongly challenge the TTI model of SU UMa star
outbursts. 

We now give evidence that the orbital hump simply results from the
orbital aspect of the superhump dissipation pattern in high
inclination systems like WZ Sge, and that the observed eclipses are
eclipses of the superhump light source itself. Thus an increased hot
spot brightness and the inferred increased mass transfer rate
are spurious, the result of a misinterpretation of observed light
curve.

We first note that the basic pattern of light curves during the common
superhump stage repeats more or less with the superhump beat period
of roughly 5 days. Eclipses were visible for only half of these beat
phases. Eclipse characteristics such as eclipse depth, eclipse width,
and time of mid-eclipse in the orbital light curve all were functions of
the beat phase (see Fig. 21 of Patterson et al. 2002).

\begin{figure}
\begin{center}
\includegraphics[width=6.8cm]{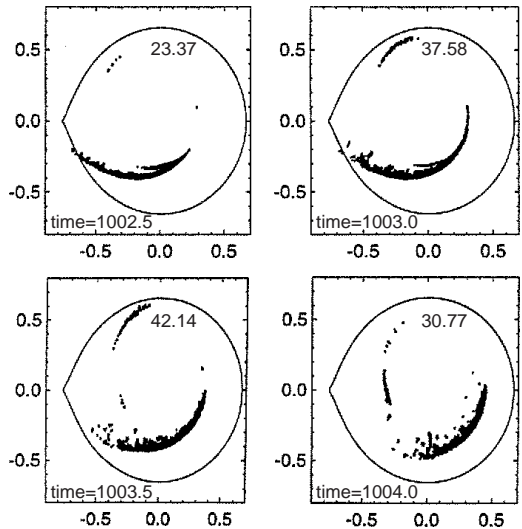}
\caption{Superhump dissipation pattern (black dots) in hydrodynamic
simulations for four different superhump phases around superhump
maximum. Axes: Cartesian coordinates in orbital plane in units of
binary separation. Contour line: primary Roche lobe; time: time
elapsed in simulation in units of orbital period /2$\pi$; number
within Roche contour: strength of dissipation (arbitrary unit). After
Murray (1999).}
\end{center}
\end{figure}

Now the superhump light variation is an intrinsic variation produced
by time varying dissipation as the secondary orbits the eccentric
disk. But the dissipation pattern is also extremely non-axisymmetric
as can be seen e.g. in hydrodynamic simulations by Hirose \& Osaki
(1990, their Fig. 7b) and Murray (1998, his Fig. 8). Fig. 1 shows the
superhump dissipation pattern for the four (out of nine) phases around
superhump maximum from Murray's Fig. 8. The most conspicuous feature
in the simulations is the appearance of an arch-like tail that
extends near to the edge of the primary Roche lobe when the
dissipation is strong (near superhump maximum). Part of this arch-like
tail can be eclipsed in the case of WZ Sge. When the tidal dissipation
is weak (superhump minimum) no conspicuous feature in the tidal
dissipation is visible and as far as the dissipative light is
concerned the disk looks compact and circular.

The tidal dissipation of the superhumps occurs near the disk rim in a
region about 1/10 of the disk radius wide (see Fig. 1). This is the
same size as the estimated half thickness of the disk at its rim. Thus
about one half of the superhump light may be expected to be radiated
from the two horizontal disk surfaces and one half from a more or less
vertical disk rim. The radiation pattern of the latter part is
strongly dependent on the orbital aspect of the observer. 

We have estimated how much light an observer near the orbital plane
would receive from this component due to the projection of its
radiating surface at the various orbital phases for all nine
superhumps phases displayed in Fig. 8 of Murray (1998). The resulting
orbital light curves are shown in Fig. 2. Note the strong variation of
the total dissipated light given by the numbers inside the individual
graphs. An intensity weighted superposition of these graphs yields an
``orbital light curve'', that corresponds to the synchronous
superposition of the superhump light curves at the orbital period by
Patterson et al. . Fig. 3
gives the resultant light curve. It shows a hump with maximum phase
around 0.75 and extending over half the orbital periods in fairly good
agreement with the observed light curve shown in Fig. 7 of Patterson
et al. (2002). The strong asymmetry of the superhump dissipation pattern thus naturally can explain the apparent orbital
hump and no increased hot spot light is needed. 

\begin{figure}
\begin{center}
\includegraphics[scale=0.75]{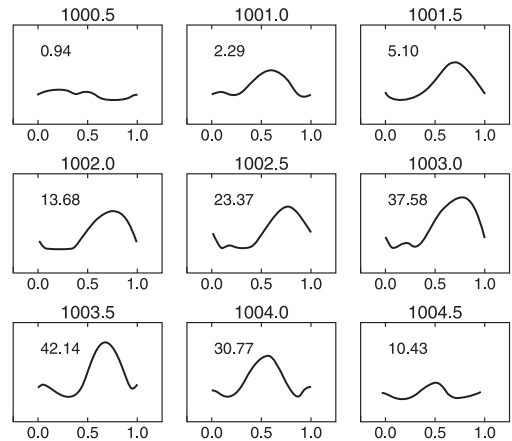}
\caption{Estimated orbital light curves from superhump dissipation
pattern projected for an observer at high inclination for 9 different
superhump phases from Murray's (1999) simulation. Horizontal axis:
orbital phase; vertical axis: intensity in arbitrary units; number on
top: time elapsed in simulation in units of orbital period  /2$\pi$;
number  inside frame: strength of dissipation (arbitrary unit).}
\end{center}
\end{figure}

\begin{figure}
\begin{center}
\includegraphics[scale=0.75]{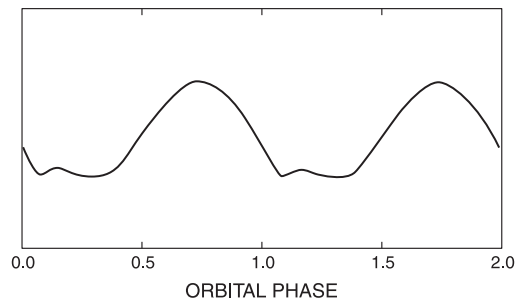}
\caption{Superposition of the nine light curves of Fig. 2 weighted by
their relative dissipation strength. Horizontal axis: orbital phase;
vertical axis: intensity (arbitrary unit). This corresponds to a
synchronous summation at the orbital period of observed superhump light
curves}
\end{center}
\end{figure}

There is even intrinsic evidence in Patterson's et al. (2002) data
that the observed orbital light curves and eclipses are indeed due to
the superhump light source and not due to a mass transfer hot spot. In
Fig. 18 Patterson et al. (2002) show the changes of eclipse depth and
orbital wave height as evidence for a changing mass transfer
rate. However these two variations closely follow the variation of the
superhump light amplitude itself, listed in their Table 3. This
indicates that it is the superhump light source which is eclipsed and
whose aspect varies with the orbital phase rather than a hot spot. 

Why are eclipses not observed in the first 12 days of the outburst
when the early humps (or ``OOH'') were seen? Osaki and Meyer (2002)
have suggested that early humps are produced by the two-armed
dissipation pattern of the 2:1 Lindblad resonance. Though at the disk
rim this spiral dissipation pattern is located azimuthally far away
from the secondary star (Fig. 3a in Lin \& Papaloizou 1979). Thus it is
fairly likely that this light is not eclipsed. 

We conclude that there is no evidence for enhanced mass transfer in
the common superhump stage of the 2001 outburst of WZ Sge.

Another case quoted by Smak (1996) as evidence for enhanced mass
transfer are eclipse observations during the superoutburst of Z Cha by
Warner and O'Donoghue (1988). O'Donoghue (1990) analyzed these
observations and found that, with the exception of one case, the
superhump light source was not confined to the hot spot region. In one
exceptional case called eclipse 77878 (Fig. 2a in Warner \&
O'Donoghue 1988), he found that all superhump light was confined to a
region near the position of the quiescent hot spot. O'Donoghue (1992)
concluded that ``as the bright spot during this eclipse is more than
twenty times as luminous as at quiescence, a brief period (due to the
rarity of this kind of eclipse) of enhanced mass transfer is the most
natural explanation''. 

However, if we recall that the superhump light source is extremely
non-axisymmetric and mostly confined to an arch-like tidal tail
extending in the following lune of the accretion disk, it is most
natural to assume that the light source eclipsed in eclipse 77878 was
just the superhump light source itself rather than the hot spot. 

In fact, as the light curve of eclipse 77878 in Warner \& O'Donoghue
(1988) shows, the eclipse occurred at the rising phase to the superhump
maximum and the shoulder of the superhump light curve was eclipsed. We
may conclude that what was eclipsed in this case is also the superhump
light source itself and not a hot spot, and that there is no evidence
for enhanced mass transfer in this case either.

\section{Theoretical aspects for enhanced mass transfer from irradiated 
secondary stars} 

In the preceding sections we have shown that all claimed
evidence for increased mass transfer in outbursts of
cataclysmic variables is inconclusive, partially
contradictory, and in all cases can be explained otherwise, in
fact supporting the TTI model in its various features.

What is the situation from the theoretical side?
The luminosity often rises in outburst by orders of magnitude and one
might expect strong irradiation of the secondary star, raising the
surface temperature and lifting gas higher up in the gravitational
potential. Why should this not lead to a density increase in the
Lagrange point $\rm{L_1}$ and an increase of the mass overflow rate?
There are however well known difficulties with this simple picture,
and we show in the following that simple quantitative estimates in the
complex physical situation yield prohibitive limits on the increase of
the mass overflow. We will perform the estimates for the case of WZ
Sge. This should be a particularly sensitive system for such effects
because of the possibly very low surface temperature of its secondary
star and an extremely high luminosity increase in the outburst.

The main problems with an irradiation effect on the mass transfer
are due to (1) the shadow that the accretion disk casts on the $\rm{L_1}$
point and (2) the high opacity in the extreme ultraviolet band which
prevents the bulk of the radiative flux from reaching the stellar
photosphere at all. In addition (3) the light that penetrates to the stellar
surface heats up the subsurface layers but this heated material cannot
effectively move towards the shadowed $\rm{L_1}$ point since strong
Coriolis forces  prevent it from moving into the lower pressure region
there.

\subsection{The shadow of the accretion disk}
Light sources are the white dwarf, the inner accretion disk and possibly a
corona above the inner disk. Important for our analysis of the effect
of irradiation is to which degree  the disk casts a shadow on the 
region around $\rm{L_1}$.

Disk structure computations allow determination of the thickness of the
accretion disk in the outburst. Near the outer radius $r_{\rm d}$ of
the disk at $10^{10.4}$ cm around a white dwarf of mass  
$M_{\rm{WD}}=0.7M_\odot$ and for an estimated accretion rate for
WZ Sge in the early phase of outburst of $10^{-7}M_\odot/{\rm {yr}}$
one obtains (Meyer \& Meyer-Hofmeister 1982) for the height of the
disk photosphere above the midplane

\begin{equation}
H_{\rm d} =10^{9.4} \rm{cm}.
\end {equation}
The height varies weakly  with the accretion rate, proportional to
$\dot M^{1/6}$. This height $H_{\rm d}$ is larger than the radius of
the white dwarf  
\begin{equation}
R_{\rm {WD}}=10^{8.9} \rm{cm}
\end {equation}
(Nauenberg 1972). It is also larger than the vertical extent of the
mass transfer cross section in the  $\rm{L_1}$ point. The latter is
obtained as the vertical scaleheight of a gas column of temperature
$T_0$ in the Roche geometry at $\rm{L_1}$ (e.g. Meyer \&
Meyer-Hofmeister 1983, Kolb \& Ritter 1990). For the assumed mass ratio $q=0.1$
and a surface temperature $T_0=10^{3.2}$K of the secondary this gives 
\begin{equation}
H_{\rm{L_1}}=10^{8.3} \rm{cm}.
\end {equation}
The value varies proportional to $T_0^{1/2}$. This means the region around
$\rm{L_1}$ is in the shadow of the disk rim.
Even light from a corona above the inner disk is shielded by the disk
rim since the height $H_{\rm d}$ of the accretion disk rim is also larger than
the scaleheight of a corona $H_{\rm c}$ above the inner accretion
disk, for a coronal temperature of $10^8$K at the distance $r=10^9$cm from
the white dwarf
\begin{equation}
H_{\rm{c}}=\left(\frac{2\frac{\Re}{\mu} T r^3}{GM}\right)^{1/2}=
10^{8.7}\rm{cm}
\end {equation}
with $\Re$ gas constant, $\mu$ molecular weight, $G$ gravitational constant.

We conclude that the region of mass transfer at $\rm{L_1}$ lies deep in
the shadow band that extends $10^{9.4}$cm to both sides of the equator
of the secondary star for any light that comes from the white dwarf,
an inner disk or a corona above the inner disk.

\subsection{The irradiated part of the surface of the secondary star}
While the mass transfer point $\rm{L_1}$ lies in the center of the
broad shadow of the accretion disk and receives no direct light the
possibility remains that irradiation raises the temperature
outside of this shadow zone and that by exchange of heated material
from the irradiated part with cool material in the shadow zone below
$\rm L_1$ the temperature also rises there. That could still result
in an increased mass transfer. We now discuss the physical aspect
of this situation in order to see whether there is an effect
on the mass transfer rate.

First we derive an estimate for the temperature of the irradiated
surface of the secondary star. In quiescence the surface receives the
radiation from the hot white dwarf, $T_{\rm{WD}}=15000\rm{K}$ (Cheng
et al. 1997). If this radiation is fully absorbed and reradiated from
the secondary surface as black body radiation its temperature at the distance
closest to the white dwarf would be 
\begin{equation}
T_{\rm{irr}}=\big(\frac{R_{\rm{WD}}}{ax_{\rm{L}}}\big)^{1/2}
\cdot T_{\rm{WD}}=10^{3.4} \rm{K}
\end{equation}
where $x_{\rm L}=r_{\rm L}/a$=0.72 is the distance of $\rm L_1$ from
the primary in units of the separation $a$, here for the assumed
mass ratio $q=0.1$.

In outburst, with the accretion rate $\dot M= 10^{-7} M_\odot/{\rm {yr}}$,
the accretion luminosity of the white dwarf will rise to the value 
\begin{equation}
L_{\rm{WD,outburst}}=\frac{1}{2}\frac{GM_{\rm{WD}} \dot M}{R_{\rm{WD}}} =
 10^{35.6} \rm{erg/s}.
\end{equation}
The corresponding effective temperature is 
\begin{equation}
T_{\rm{WD,outburst}}=\big(\frac{L_{\rm{WD}}}
{4 \pi \sigma R_{\rm{WD}}^2}\big)^{1/4}
 = 170000\rm{K}.
\end{equation}

Radiation of this temperature lies in the extreme ultraviolet and
supersoft X-ray range with wavelengths around 200 \AA \,  where
opacities are very large. Such radiation cannot reach the surface of
the secondary star but is absorbed and reradiated in high layers of the
secondary's atmosphere (see Suleimanov et al. 1999 for the analogous
case of supersoft X-ray sources for a detailed
discussion). Hard X-ray flux above 2keV on the other hand could
penetrate to the photosphere. But RXTE observations between 2 and 6
keV yielded upper limits of $f_{\rm x}=5.3\cdot 10^{-12}$ and $3.7 \cdot
10^{-12} \rm{erg/cm^2s}$ (for the flux density at earth) on the
day of discovery and 4 days into the last outburst of WZ Sge (Kuulkers
et al. 2002). With a distance to WZ Sge of about $d$=50 pc (Smak 1993,
Meyer-Hofmeister et al. 1998) this corresponds to upper limits for
the X-ray luminosity

\begin{equation}
L_{\rm{x}}=4\pi d^2 f_{\rm x} \leq 10^{30.2} \rm{erg/s},
\end{equation}
an order of magnitude less than even the luminosity of $10^{31.4}
\rm{erg/s}$ of the 15000 K quiescent white dwarf.
Contrary to the extreme UV, near UV and optical radiation from the
accretion disk and white dwarf however can reach the photosphere of
the secondary and raise its temperature. From HST observations of WZ
Sge (Kuulkers et al. 2002) 8 days into the 2001 outburst we estimate a
flux density longward of $\lambda$=2000\AA \ of
$f_{\rm{UV}}{'}=10^{-8.5}\rm{erg/cm^2s}$ at earth. For a distance of 50
pc this corresponds to a luminosity
\begin{equation}
L_{\rm{UV}}{'}=4\pi d^2 f_{\rm{UV}}{'}= 10^{33.0} \rm{erg/s}.
\end{equation} 
If this radiation is reprocessed at the secondary's
photosphere it raises the surface temperature to
\begin{equation}
T_{\rm{irr}}=\big(\frac{L_{\rm{UV}}{'}}
{4\pi (ax_{\rm{L}})^2 \sigma}\big)^{1/4}
= 10^{3.8} \rm{K}
\end{equation}
at the distance $ax_{\rm L}$ of $\rm L_1$ from the white dwarf.
This is a maximal
value. Surface elements farther away from the central source and inclined
to the direction of irradiation will have a lower surface temperature.

It is worth noting that this temperature estimate of
$10^{3.8}$K is significantly lower than the value of $10^{4.32}$K that
Smak (1993) derived from spectral fits to IUE observations early
in the December 1978 outburst of WZ Sge. He attributed this component
to the irradiated surface of the secondary star but noted an
unexplained inconsistency between the observed and expected phases of
this component in the light curve.

Our significantly lower estimate for the surface temperature of the
secondary star in outburst provides additional support
that the source of the component determined by Smak is not the
surface of the irradiated secondary but is the 2:1 resonance
dissipation pattern at the rim of the accretion disk as suggested 
in Osaki \& Meyer (2002). For that dissipation pattern a temperature
around $10^{4.3}$K was estimated.
The heat deposited by irradiation at the surface of the secondary
spreads with time by radiative diffusion into the subsurface
layers. The depth down to which the temperature is raised depends on
the opacity of the gas and the duration $t$ of the irradiation. 
Using standard radiative diffusion and opacities in the temperature
range between $10^{3.2}$K and $10^{3.8}$K taken from Alexander et al. (1983) 
we estimate the column density of the affected layer as
\begin{equation}
\Sigma=10^{2.8}\big({\frac{t}{3^{\rm d}}}\big)^{1/2} \rm{g/cm^2}. 
\end{equation}
This is smaller than the column density $\Sigma=10^{3.9} \rm{g/cm^2}$
down to the level where the unirradiated star of surface temperature
$10^{3.2}$K itself reaches the $10^{3.8}$K of the irradiated surface.
For this estimate we used a value of $d\ln T/d\ln P
=0.15$, a mean value near the surface of very cool low mass stars
(B\H{u}ning, private communication) and a pressure at the surface
$P_{\rm s}$ of $10^{3.7} \rm{dyn/cm^2}$ (which for WZ Sge corresponds 
to a mass overflow rate of $10^{-11}M_\odot/{\rm{yr}}$).

Thus below the irradiated surface of the secondary a more or less
isothermal layer of $T=10^{3.8}$K is formed.
\subsection{Heat transport from irradiated to shadowed regions?}
The heating of the subsurface material under the irradiated surface
raises the pressure and causes pressure differences on equipotential
surfaces between irradiated and unirradiated parts. They tend to drive
heated material into the shadow belt and below $\rm{L_1}$ which could
raise the pressure there and lead to an increase of the mass transfer
rate.

However, the system is rotating with the binary period and due to
deformation of the stellar surface near $\rm{L_1}$ the rotation
frequency has a component normal to the surface. For q=0.1 along the
meridian through $\rm{L_1}$ one obtains for this normal component 
\begin {equation}
\Omega_\perp =\Omega_{\rm {orb}}\cdot \cos \vartheta=
\,
 0.56 \Omega_{\rm {orb}}
\end {equation}
where $\Omega_{\rm {orb}}=2\pi / P_{\rm {orb}}=10^{-2.9}\rm{s^{-1}}$
is the orbital angular frequency ($P_{\rm {orb}}$ orbital period) and
$\vartheta$ is the angle between the surface normal and the rotation axis
of the system. In the Roche geometry this angle changes gradually as
one moves away from the meridian. But along the meridian is the
shortest path out of the shadow band and the driving pressure gradient
is the largest in this direction. We will show now that rotation
practically prevents heated matter to flow into the low pressure
region below $\rm{L_1}$.
For this we simplify the curved shape of the equipotential surfaces
near $\rm{L_1}$ by adopting a locally plane geometry with Euclidian
coordinates $x$ in the direction of constant latitude, $y$ in the
direction of constant longitude, and $z$ in the direction of the outward
surface normal.

The horizontal equations of motion in this system are used by
Spruit (2002) in the context of a solar problem. There it is shown that
the equations can be simplified if the vertical extent is
 small
compared to the latitudinal extent which is the case here with a few
vertical pressure scaleheights of order $10^8$cm, small compared to
the width of the shadow band of $10^{9.4}$cm. Since the main gradients
are perpendicular to the shadow band which spreads along the equator we
neglect gradients in the $x$ direction and obtain (Pedlosky 1982, Spruit 2002)
\begin {eqnarray}
-\frac{\partial P}{\partial y} -2\rho v_x\Omega_\perp
 +\frac{\partial}{\partial z} \big(\rho\nu
 \frac{\partial v_y}{\partial z}\big) &=& 0,
\nonumber \\
2\rho v_y \Omega_\perp  
 +\frac{\partial}{\partial z} 
 \big(\rho\nu \frac{\partial v_x}{\partial z}\big) &=& 0.
\end {eqnarray}
$\nu$ is the kinematic viscosity. Without viscosity we have the
``geostrophic'' flow along the shadow band
\begin {eqnarray}
v_x &= & -\frac{1}{2\Omega_\perp} \frac{1}{\rho} \frac{\partial P}
{\partial y} \approx 
-\frac{1}{2\Omega_\perp}
\frac{\Delta V_{\rm s}^2}{l}
\nonumber \\
v_y &= & 0
\end {eqnarray} 
where $V_{\rm s}=(P/ \rho)^{1/2}$ is the isothermal sound velocity and
$l$ is the half width of the shadow band. Thus the flow velocity is at right angle to the driving pressure gradient which is balanced by the
Coriolis force, similar to flows in weather systems and oceans on the
rotating earth. Thus without friction no heat can be advected into
the shadow region below $\rm{L_1}$.

If a small viscosity is present the geostrophic flow is accompanied
by a small drift into the low pressure shadow band. Replacing $z$
derivatives by division by the vertical scale height $H$ in the
equation for $v_y$ one obtains 
\begin{equation}
v_y= \frac{\nu}{2\Omega_\perp H^2}v_x =
-\frac{\nu \Delta V_{\rm s}^2}{4\Omega_\perp^2H^2 l}.
\end{equation}
What is the value of a possible viscosity? Molecular viscosity is
negligibly small. Can there be turbulent viscosity? The layer
irradiated and heated from above is stable against convection
as the temperature
is constant or even increasing upwards. This suggests that there will
be practically no viscosity in the main body of the heated
material. Only at the bottom (i.e. its densest layer) one might have
to consider that convection from the original stellar envelope below
penetrates into the heated layer above. We will now estimate the
effect that this will have on the advection of heat into the shadow
zone. For this we determine from mixing length theory
a ``convective'' viscosity
\begin{equation}
\nu=\frac{1}{3} l_{\rm m}v
\end{equation}
where $l_{\rm m}$ is the mixing length, e.g. the pressure
scaleheight $H$, and $v$ is the convective velocity of the cool
secondary at the depth where the overlying irradiation heated zone
starts. The velocity $v$ follows from the requirement that $v$ transports the
convective flux, which is equal to the flux $\sigma {T_0}^4$ radiated
away at the surface of the unirradiated secondary. Using the value of
$d\ln T/d\ln P = 0.15$ to determine the specific heat and
for a density $\rho=\Sigma /H =10^{-4.2} \rm{g/cm^3}$
($\Sigma =10^{2.8} \rm{g/cm^3}$, Eq.(11), and $H=10^7 \rm{cm}$) one
obtains (Kippenhahn \& Weigert 1990, Eqs. (7.6) and (7.7)) $v=10^{3.5}
\rm{cm/s}$. The value for $H$ results from the surface gravity in Roche
geometry at the distance $l=10^{9.4}$cm of the shadow band boundary from
the Lagrange point $\rm {L_1}$ and for a temperature of $10^{3.8}$K.
Eq. (16) then gives 
\begin{equation}
\nu=10^{10}\rm {cm^2/s}.
\end{equation}
The pressure ratio between surface, $P_{\rm s} =10^{3.7}\rm{dyn/cm^2}$,
and the bottom layer, $P= \frac{\Re}{\mu} T
\rho=10^{7.4} \rm{dyn/cm^2}$, determines the temperature ratio between
surface and bottom and with the surface temperature of the
unirradiated star $T_0=10^{3.2}$K this gives
$T=T_0{(P/P_{\rm s})}^{d\ln T/d\ln P}=10^{3.75}$K, already very close to the temperature
of the irradiated layer of $10^{3.8}$K.

With the value for $\nu$ and $\Delta V_{\rm s}^2=0.1 V_{\rm s}^2=
10^{10.6}\rm{cm^2/s^2}$ the drift velocity into the shadow zone
becomes 
\begin{equation}
|v_y| =10^{2.9}\rm {cm/s}.
\end{equation}

This means that the time for matter to drift from irradiated parts into the
region below ${\rm L_1}$ becomes 
\begin{equation}
t_{\rm{drift}}=\frac{l}{|v_y|} =10^{6.5} \rm s = 37^{\rm d}
\end{equation}

This is longer than the duration of the outburst. It is also much longer
than the cooling time of three days of the column considered.
The matter therefore has cooled to temperatures in the shadow zone
long before it reaches the region below $\rm{L_1}$.

Thus the shadowing  of the mass transfer region by the accretion disk
rim and the strong Coriolis force prevent any effect of irradiation of
the secondary star on the mass transfer. This conclusion agrees with
the observational evidence that is discussed in the preceding
sections.

We discussed this here for the case of WZ Sge
in particular, but we expect the same effect to be true for other CV
systems as well.

\section {Smak's (1996) criticism and a new picture of the TTI model }
 
Smak (1996, 2000) argued that ``the TTI model faces a serious problem 
with the sequence of events it predicts:  the superhumps should appear 
at an early phase of a superoutburst (certainly not later than 
its maximum), while observations show that this happens only one 
or two days {\it after} maximum". In the original TTI model, the last normal 
outburst triggers a tidal instability, leading to an eccentric disk
which causes a major enhancement of the accretion rate. Accordingly, the superhumps should 
appear in the very early phase of a superoutburst. 
Therefor Smak (1996) suggested 
instead a hybrid model (called EMT model) by combining the tidal 
instability with the enhanced mass-transfer model. 
In Smak's picture,  irradiation 
of the secondary star by the central white dwarf and the boundary 
layer during the superoutburst causes enhanced mass transfer which 
 now can keep the disk in hot state for much longer time.  
There is then enough time for the disk to develop the eccentric 
structure, a necessary condition as claimed by Whitehurst and King (1991).

In response to Smak's criticism, we here propose a refinement 
 of the TTI model. 
Before  doing so, let us first examine the observations more carefully. 
Bateson (1977) classified the superoutburst light curves of VW Hyi 
into types S1 to S8  but these are basically two types; S1-S5 with simple 
superoutburst light curves with one continuous rise and fall   
and S6-S8 with light curves of a precursor-main outburst type.  
Almost all observers note that the rising part of the superoutburst 
is indistinguishable from that of the normal outburst. 
As discussed by Marino and Walker (1979), in the case of the S6-S8 types the precursor normal outburst 
was separated by a dip from the main part of the superoutburst 
 while it is already merged with 
the main superoutburst in the case of S1-S5. That is, in the type S1-S5 
superoutbursts, the time delay between the triggering normal outburst and 
the bulk of the superoutburst is too short to be observable (Marino \& 
Walker 1979; van der Woerd \& van Paradijs 1987).  

There are several examples in which the superoutburst light curve in 
optical light looked more or less as one single continuation but 
it consisted of a  precursor and a main outburst at shorter 
wavelengths.  In general, precursor 
part and main part of the superoutburst are more clearly separated 
as we go to shorter wavelengths. This can be seen in the Voyager 
far-ultraviolet observations of the 1984 October-November superoutburst 
of VW Hyi (Pringle et al. 1987), and in EUV observations of 
the 1997 March superoutburst of OY Car by Mauche and Raymond (2000)
and the 1997 October superoutburst of SW UMa by Burleigh et al. (2001).
When the precursor is well separated from the main superoutburst, the main 
outburst grows in amplitude with the growth of the superhump light.  
The long duration of the superoutburst can not be the cause of the eccentric 
disk because the superhumps grow together with the rise to the main outburst. 
As far as superoutbursts of the precursor-main outburst type are concerned, 
observations agree very well with the TTI model. 

We now turn to the S1-S5 type superoutbursts in which 
light curves look like a single continuation and the superhumps appear only 
one or two days {\it after} maximum and therefore Smak's criticism seems 
relevant. To deal with it, we here propose a refinement 
of the original TTI model  that explains why superhumps appear 
a few days {\it after} the superoutburst maximum in some SU UMa stars. 
We suggest that these are those outbursts in which the last normal 
outburst occurs in full readiness for the superoutburst and that 
the normal outburst and the superoutburst are completely merged 
as explained below. 

We first discuss the outburst duration 
in dwarf novae. In particular, let us 
address why in the TTI model the disk in normal outbursts 
can not stay hot even when much mass has already been stored 
in the disk for the coming superoutburst. The essential point of 
the TTI model lies in the explanation of the shortness of the 
normal outburst. When the normal outburst occurs, the disk jumps from the 
cold state to the hot state. The surface density distribution then 
changes from  $\Sigma\propto r$ in the cold state  to $\Sigma\propto
r^{-0.75}$  in the hot state. 
On the other hand, the critical surface density 
below which no hot state exists varies with the radius as 
$\Sigma_{\rm crit}\propto r$.   Thus the outer rim of a free accretion 
disk (by ``free" we mean not yet bound by the tidal removal 
of angular momentum) will always peter out to surface densities 
below the critical density below which no hot state is
possible. When the heating wave released in the outburst reaches this
critical surface density
a cooling wave is set up that moves inward and cuts off the (normal) outburst. 
In the TTI model, the normal outburst is an outburst in which the 
disk's outer edge does not yet reach the 3:1 resonance radius and thus we
have a rather ineffective tidal removal of angular momentum 
from the disk, explaining its short duration. 

On the other hand, if tidal removal of angular momentum is very effective, 
the expansion of the disk is stopped, i.e., the disk hits a tidal wall, 
and the matter is dammed up there. If there is enough mass in the disk,
the surface density at the tidal wall is above the critical surface
density and the disk 
stays in the hot state, and the long viscous depletion of mass 
from the disk ensues. This explains the long duration of the outburst. 
 In the case of U Gem stars the tidal wall is provided at the ordinary 
tidal truncation radius, in the case of SU UMa stars 
at the 3:1 resonance radius, and in the case of WZ Sge at the 2:1 resonance. 

Let us now come back to the problem of S1-S5 type superoutbursts. 
In SU UMa stars with relatively long orbital periods, near two hours, 
having binary mass ratios around 0.2 such as VW Hyi, 
the tidal truncation radius lies just beyond the 3:1 resonance radius. 
If the final normal outburst occurs in full readiness, 
the disk will not only expand to 
the 3:1 resonance radius but also pass it. The disk will then reach 
the tidal truncation radius so that its expansion is stopped there 
(see, the upper right-most panel of Fig. 4). 
The viscous depletion of matter (i.e., the viscous plateau stage of 
outburst) ensues even when the eccentric tidal instability 
(i.e., superhumps) has not yet developed to observable amplitude. 
One or two days after the light maximum, the tidal instability 
and the superhumps have grown to large amplitude, now taking over 
the role of tidal removal of angular momentum from the ordinary 
tidal torques at the truncation radius.  
This may be the reason why in these outbursts the superhumps grow one
or two  days {\it after} the maximum. 

\begin{figure}
\begin{center}
\includegraphics[scale=0.75]{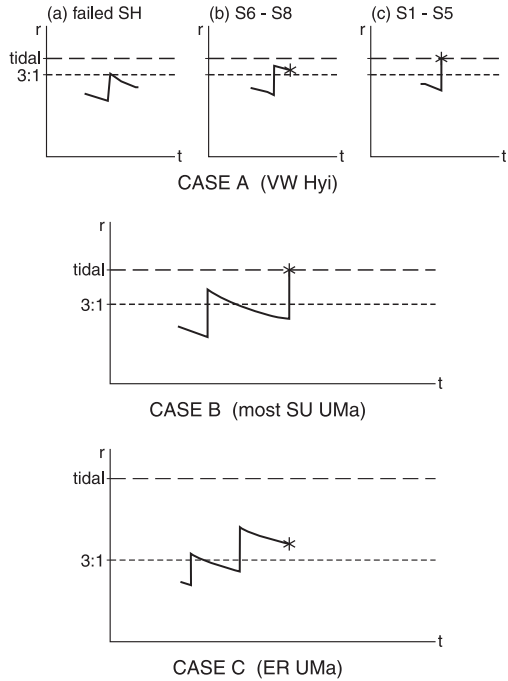}
\caption{Various ways in which superoutbursts are triggered (see text
for details). Horizontal axis: time, vertical axis: radius. Short
dashed line: 3:1 resonance radius, long-dashed line; tidal truncation
radius, solid line: outer disk radius, asterisk: ignition of
superoutburst. In order of decreasing mass ratio (increasing
separation of tidal and 3:1 resonance radii, increasingly longer
growth time of superhumps): case A: relatively large mass ratio around
0.2 (a), (b), and (c): ordinary outbursts occurring in various stage
of readiness for a superoutburst in one and the same system. Case B:
smaller mass ratio. Case C: extremely small mass ratio.} 
\end{center}
\end{figure}

In the original suggestion of the TTI model (Osaki 1989, 1996), 
the importance of the tidal truncation radius just beyond the 3:1 resonance 
radius was not appreciated and only the precursor-main outburst type  
superoutburst was dealt with. With this new refinement we can now understand different aspects of superoutbursts of 
various SU UMa stars. 
Fig. 4 illustrates how the last normal outbursts trigger the superoutburst 
under different conditions. Each panel schematically 
shows the time variation of the disk's outer edge. The asterisk denotes
the point where the superoutburst is triggered.   
The uppermost three panels (case A) represent  the longest 
orbital-period SU UMa stars with  periods around two hours 
(typified by VW Hyi) and binary mass ratios $q$ around  0.2 
for which the 3:1 radius and the tidal truncation radius are close to 
each other. The upper left panel illustrates an aborted superhump 
during the last normal outburst, which fails to develop to the 
superoutburst discussed in subsection 2.2.  
The upper middle panel illustrates the case in which the last normal 
outburst first triggers a tidal instability and then the superoutburst 
develops together with the superhumps, corresponding to the 
precursor-main outburst type of S6-S8. The upper right panel illustrates 
the case of full-readiness in which the disk expands to reach  
the tidal truncation radius and the hot viscous plateau stage sets in.
The tidal instability and the superhumps then develop to large amplitudes, 
corresponding to the S1-S5 case just discussed above. 

 The middle panel (case B) shows SU UMa stars with shorter orbital
periods that have lower q values. 
In this case, even if the outer edge of the disk reaches the 3:1 resonance, 
the growth rate of the tidal instability, proportional to $q^2$, is so
low that  there is not 
enough time for the superhump to grow to large amplitude 
within the normal outburst duration. Once the normal outburst ends 
and the disk returns to quiescence, the outer edge of the disk 
contracts due to the addition of matter with low specific angular
momentum and shrinks below the 3:1 resonance.
Observationally, the normal outburst in this case is indistinguishable  
from those normal outbursts in which the disk's edge does not 
reach the 3:1 resonance at all. A next normal 
outburst will then push the disk edge to the tidal truncation radius. 
The hot viscous plateau stage sets in. The superhumps develop  
a few days {\it after} the light maximum. Most SU UMa stars correspond 
to this case B. 

The bottom panel (case C) illustrates the case of the shortest orbital 
period SU UMa stars with the lowest mass ratio $q$ in which normal outburst 
occurs frequently. In this case the tidal truncation radius is far away from 
the 3:1 resonance radius. As discussed for case B, the normal outburst, 
in which the disk's outer edge reaches the 3:1 resonance, can not develop 
the superhumps to large amplitude. This may repeat in the next 
normal outburst as well. However, since the disk expands more and more
above the 3:1 
resonance radius, it may finally not shrink below the 3:1 resonance radius 
even during quiescence and the tidal instability continues to operate 
in quiescence. If sufficient time elapses, the tidal instability 
and the superhumps grow to large amplitude. Tidal heating at the outer edge 
of the disk will then force the upward thermal transition 
to the hot state, starting a new outburst. This outburst is of outside-in type 
and it develops always to a superoutburst because the disk is now fully 
eccentric, that is, a superoutburst triggered not by the normal outburst but 
rather by the tidal instability. 

An obvious prediction in this case is that the superhump should be visible 
in large amplitude even during the rising branch of the outburst. 
This case (case C) corresponds to ER UMa stars with the shortest 
supercycle less than 50 days, and with the very short normal-outburst 
cycle, as short as 4 days. (The WZ Sge stars do not belong to this 
case because normal outbursts do not occur so frequently). It was a puzzle 
in observations of the ER UMa stars that superhumps in large amplitude 
were observed during the rising branch of the superoutburst (see  observation 
of ER UMa  by Kato et al. (1996) and V1159 Ori by Patterson et al. (1995)). 
In the standard scenario of SU UMa stars, the superhumps were expected 
to appear later during the superoutburst  the lower the mass 
ratio of the system is because of their increasingly longer growth
time. The mystery of these paradoxical 
observations is now solved.    

Although the 3:1 resonance and the ordinary tidal truncation both 
remove angular momentum from the disk via the secondary's tidal force, 
there exists an essential difference in nature between them. 
In a sense  the 3:1 resonance is more active but 
the ordinary tidal truncation is more passive in nature: 
When the eccentric disk (and the superhump) is fully developed 
by the 3:1 resonance, tidal removal of angular momentum from the disk 
is very strong as demonstrated by Murray (1998) with SPH hydrodynamic 
simulation. This keeps the disk in the hot state longer. When the superoutburst ends, 
the quiescent disk becomes quite compact and far removed from the 3:1
resonance 
radius. On the other hand, the ordinary tidal truncation is passive 
in that it simply inhibits the disk to expand beyond the truncation 
radius but it does not cause the disk to contract far below that limit.    

We conclude that the TTI model fully explains the sequence of events 
of SU UMa stars and that Smak's criticism is properly dealt with 
by the new refinement.

\section{The overall picture of the 2001 outburst of WZ Sge}

We now give a description of the overall development of the 2001 
outburst of WZ Sge in terms of the TTI model, based on a very low
viscosity in quiescence and no enhanced mass transfer during the
outburst. In this latter point we differ from the observational
account by Patterson et al. (2002). As shown in Sect. 2.3 their
evidence for increased mass transfer is probably the result of a
misinterpretation of orbital light curves in the common superhump era.

\noindent (1) {\it The main accretion event.}

Osaki (1995) has already 
pointed out that the basic outburst energetics of WZ Sge fits to Osaki's 
(1974) disk instability model in its simplest form: long quiescence of 
about 30 yrs coupled with a low mass transfer rate $\sim 10^{15}$gs$^{-1}$  
allows the disk to accumulate mass $\sim 10^{24}$g, which is 
dumped onto the central white dwarf during an outburst. 
Cannizzo (2001) has shown that the 2001 superoutburst 
of WZ Sge can be explained by viscous depletion of matter in a disk 
with an initial mass $\sim 10^{24}$g. 
\\

\noindent (2) {\it Quiescence and long recurrence time.} \\
Two different models have been proposed to explain the long recurrence
time of WZ Sge of about 30 yrs. One is a model with extremely low viscosity 
in quiescence (Smak 1993; Osaki 1995, Meyer-Hofmeister et al. 1998). 
The other is a model of a steady 
or almost steady cold disk with a standard value of viscosity 
in the cold state but with its inner part truncated 
(Warner et al. 1996; Hameury et al. 1997). 
Such a disk is unable to store the amount of mass required for the
outburst. (See Fig. 3 of Meyer-Hofmeister et al. 1998 for a comparison
of the two models).

The missing mass in the second model would have to be supplied during
the outburst. If $10^{24}$g would be transferred on a time scale of
one day  the enhanced mass transfer rate should be 
$\dot M_{\rm tr} \simeq 10^{19}$gs$^{-1}$ and it may be rather unlikely to 
overlook such a large increase. As discussed in 
Sect. 2.3, there is no good observational evidence 
for enhanced mass transfer during the outburst.
  
The very low quiescent viscosity in the first model can be understood
by decay of magnetic turbulence (Gammie \& Menou 1998). In WZ Sge this
occurs during the final decline to quiescence (Osaki et al. 2001). The
extremely low viscosity value of WZ Sge stars may result as consequence
of a cool brown dwarf secondary that cannot support a magnetosphere
whose interaction with the accretion disk has been made responsible
for the standard quiescent viscosity of ordinary dwarf novae (Meyer \&
Meyer-Hofmeister 1999). 
\\

\noindent (3) \it{Start of outburst and early humps.} \rm{}\\
Least understood in this scenario is the initial kick to start the 
thermal instability in the case of extremely low quiescent viscosity. 
Once the thermal instability is started and the disk goes into the hot state, 
the standard high viscosity is most likely produced by the magneto-
rotational instability (Balbus and Hawley instability). 

A sudden jump to high viscosity makes the accumulated mass spread
inwards and outwards and form a standard hot accretion disk. 
As discussed by Osaki and Meyer (2002), the outer edge of the
disk not only reaches the 3:1 resonance 
but passes it and reaches the 2:1 resonance radius where
the two armed spiral pattern of the Lindblad resonance is excited. Increased tidal torques 
now stop the expansion of the disk and the tidal dissipation produces 
the observed photometric double humps (early humps, or ``OOH" in Patterson 
et al. 2002).

\noindent (4) \it{Late appearance of common superhumps} \rm{} \\
Within the first 12 days the large amount of mass residing 
near the 2:1 resonance region is gradually cleared by tidal loss of angular
momentum. This
enhances the relative importance of mass around the 3:1 resonance 
(responsible for the eccentric tidal instability). 
This now allows the disk to 
develop the common superhumps as discussed below. 

Lubow (1991) has shown that an eccentric ``corotation" resonance
(different from the familiar eccentric Lindlad resonance)  occurs at the 
2:1 resonance through the tidal potential component $m=2$, which acts to 
damp eccentricity, in competition to the excitation 
of eccentricity at the 3:1 resonance. 
As long as a large amount of matter resides at the 2:1 resonance 
 the damping effect  will dominate, preventing the growth of eccentricity. 
As matter is cleared from the 2:1 resonance region, the 
excitation effect at the 3:1 resonance will increase in relative importance 
and eventually win. 
The disk eccentricity will grow to large amplitude and superhumps appear.

The tidal removal of angular momentum 
by the eccentric disk now takes over the role of the 2:1 resonance. 
The common superhump era corresponds to this stage. Thus the 2:1 resonance not only offers an explanation for the early hump 
phenomenon but it may also act as a suppressor of growth of the eccentricity 
in WZ Sge stars. The rather late appearance of the common 
superhumps in WZ Sge is then a natural consequence of the 2:1 resonance. 

Light curves in the common superhump era were already discussed 
in subsection 2.3. Orbital humps and eclipses in this stage are 
due to the superhump light source itself and there is no evidence 
for enhanced mass transfer. 
\\

\noindent (5) \it{Late superhumps.} \rm{} \\ 
The superhump light signal often stays on after the end of the main 
superoutburst of SU UMa stars and the eccentric disk seems to survive for 
a while. Light variations with the superhump period after the end of 
the main superoutburst are called ``late superhumps". 
The late-superhump phenomenon is conspicuous in WZ Sge stars, 
late superhumps continued to exist for one hundred days in the 2001 
outburst of WZ Sge.  

This phenomenon is most naturally explained by decoupling of the tidal 
instability from the thermal instability proposed by Hellier (2001). 
He suggested that in a binary system with extremely small mass ratio 
the tidal torques on the eccentric disk are not strong enough 
to keep the disk in the hot state though the tidal eccentric instability still 
goes on. The premature shutdown of the thermal instability ends 
the superoutburst even when the disk still has its eccentric form. 

When near the transition into quiescence the superhump signal has
dwindled away and the ordinary hot spot becomes dominant but the disk
still retains a residual eccentricity a modulation in phase and
intensity of the hot spot light occurs as the impact point of the
accretion stream travels around the elliptically deformed circumference
of the disk together with the revolving mass supplying secondary. This
was documented by Rolfe et al. (2001) for the eclipsing  dwarf nova IY
UMa in the very late superhump phase.
\\

\noindent (6) \it{Dip and Echo outbursts.} \rm{} \\
The WZ Sge stars often show rebrightening or reflare after the end of the 
main outburst, the most spectacular case was six consecutive rebrightenings 
of EG Cnc in 1996/1997. 
The 2001 outburst of WZ Sge showed the same phenomenon 
but it had as many as 12 short outbursts with a mean repetition 
time as short as 2 days. One of the important aspects during the echo
outburst era 
is that the superhump light signal continues to exist throughout this period 
even to the final decline stage and that the superhump amplitude stayed 
approximately constant {\it in intensity} while the background 
light intensity was highly variable.     

Osaki et al. (2001) and Hellier (2001) proposed that echo outbursts of EG Cnc 
are normal outbursts in the disk where tidal removal of angular
momentum  from the
mass reservoir beyond the 3:1 resonance radius feeds mass into the
inner  disk and repeatedly excites the thermal instability. Note that
no  enhanced mass transfer is needed. 
In particular, Osaki et al. (2001) demonstrated that this model can explain 
the sudden cessation of echo outbursts as due to decay of magnetic 
viscosity in the disk.  This model can be applied to the 2001 outburst 
of WZ Sge as well. The only difference being the amount of mass 
in the reservoir, larger in WZ Sge than in EG Cnc. 

 On the other hand, Hameury et al. (2000) have proposed enhanced mass transfer 
to explain echo outbursts. Patterson et al. (2002) supported this model 
because they saw observational evidence for enhanced mass transfer. 
However, as discussed in Sect. 2.3, their conclusion for enhanced mass 
transfer is simply due to a misinterpretation of their observations. There is 
no evidence for enhanced mass transfer and the supporting evidence 
disappears. 

Enhanced mass transfer also tends to quench an existing eccentric
structure (see Ichikawa et al. 1993; Lubow 1994) since addition of
a large amount of mass  with low specific angular momentum leads to
disk shrinkage. This would contradict the observed long endurance 
of late superhumps in WZ Sge and EG Cnc which may result just because
of the low mass transfer rate in these systems  compared to ordinary
SU UMa  stars.  
\\

\noindent (7) {\it Final decline.} \\
As discussed by Osaki et al. (2001),  the final decline 
is understood in terms of decay of magnetic turbulence in the cold disk. 
WZ Sge has entered into dormancy for another 20 or 30 yrs accumulating 
mass in the disk for the next outburst.  

\section{Conclusion}

We have carefully analyzed old and new claims of evidence for increased
mass overflow in outbursts of SU UMa stars. We find that all this
evidence is not well founded, inconclusive, and partially inconsistent
with the observations themselves. Our theoretical analysis supports
the conclusion of no increased mass overflow during the outbursts. In
all cases we find that a refined thermal-tidal instability model with
no increased mass overflow gives a consistent and convincing
explanation of the observational results. This appears to rule out
theoretical models for SU UMa star outbursts that postulate increased
mass overflow during outbursts, and allows a deeper insight into the
nature of these outbursts.
 
\begin{acknowledgements}
We would like to thank Emmi Meyer-Hofmeister for helpful discussions 
and technical assistance. 
Yoji Osaki acknowledges financial support from the Japanese Ministry of 
Education, Culture, Sports, Science and Technology with a Grant-in Aid 
for Scientific Research No. 12640237.  
We thank the referee for suggesting figures and solliciting comments
on other SU UMa stars which significantly improved our presentation of
the paper.

\end{acknowledgements}

\begin{appendix}

\section{Comments on Kato's (2002) model for early humps.}

In the context of the general outburst model for WZ Sge
we comment here on Kato's (2002) model for the early
humps of the 2001 outburst of WZ Sge, an alternative model to ours.
During the early stage of 
the 2001 outburst of WZ Sge, a two-armed arch-like pattern was observed 
in the Doppler maps of emission lines of He II 4686 $\rm{\AA}$ and C III/N III 
on the second day of the outburst, July 24,  (Baba et al. 2002) and 
on July 28 (Steeghs et al. 2001a).  Kato (2002) has proposed that this 
and the early hump 
phenomenon can be understood in terms of irradiation 
of the disk by the central radiation source. In this picture, the 
irradiation occurs preferably at the elevated surface of the disk, 
which in turn is produced by vertical tidal deformation (Smak 2001; 
Ogilvie 2002).
Thus the main energy source of early humps in Kato's model is
the irradiation of a tidally elevated disk while it is the tidal
dissipation 
in our model. The elevated disk is produced in his model 
by the vertical resonance discussed by Ogilvie (2002) while the 2:1 
Lindblad resonance is responsible for the enhanced tidal dissipation 
in our model.  
 
Here we comment on his interpretation. In our previous paper (Osaki 
and Meyer 2002), we have only dealt with photometric humps of the 2001 
outburst of WZ Sge and we have not discussed its Doppler maps. 
Here we present our view on the two-armed arch-like pattern in Doppler 
maps and its relation to the photometric humps.  
In our picture, the photometric humps are thought to be produced 
by the two-armed spiral pattern of the tidal dissipation due to the 2:1 
Lindblad resonance. The enhanced tidal dissipation may then produce a  
vertical thickening of the accretion disk in a similar azimuthal
two-armed pattern. Emission lines observed during the outburst of dwarf novae, 
in particular, those of He II at 4686 $\rm{\AA}$ and C III/N III complex 
at 4640 $\rm{\AA}$ observed on July 24 and on July 28 in the 2001 outburst of 
WZ Sge are most likely produced by the irradiation of the vertically elevated 
disk by hard photons (EUV and soft and hard X-rays) from the central source, 
in the high atmosphere of the accretion disk either by direct ionization 
and recombination or by chromospheric emission due to the temperature 
inversion as discussed by Smak (1991). We agree in this sense 
with Kato (2002) in that emission lines of He II 4686 $\rm{\AA}$ and C III/N III 
are produced by irradiation of the accretion disk by hard photons 
from the central radiation source. However, we do not agree 
with Kato (2002) on the explanation of the early hump phenomenon. 
Although two-armed arch-like pattern in the Doppler maps 
and the double-humped light curve (i.e., early humps) may be somehow 
related with each other, it is not so obvious that these two phenomena 
are produced by the one and the same cause as suggested by Kato (2002) 
because emission mechanisms are not necessarily the same for emission 
lines and for the continuum radiation. It seems premature to make a
one to one correspondence between photometric humps and features 
in Doppler maps. 

In fact, there is observational evidence that the light source 
for photometric humps is not due to the irradiation of the disk but rather 
it is due to the tidal dissipation. The observations of 
the very first day of the outburst of WZ Sge (July 23, 2001) tell us  
this. On the day of discovery of the 2001 outburst of WZ Sge, 
Ishioka et al. (2002) had found the growing humps during the rapid
rise to  maximum. On the same day, CCD spectroscopic observations 
were performed by K. Ayani at Bisei observatory, as shown in Fig. 1 of 
Baba et al. (2002). It showed Balmer lines in absorption but no emission lines 
were found around the wavelengths of He II and the C III/N III complex. 
This indicates that the irradiation of the accretion disk by hard 
photons from the central part of the disk or from the central white dwarf 
has not yet started on the first day. This is easily understood 
in dwarf nova outbursts. On the very first day, the outer part of 
the accretion disk of WZ Sge was already in the hot and high-viscosity state 
and the disk was expanding both inward and outward. The outer edge of
the disk reached the 2:1 resonance radius, 
thus triggering the Lindblad resonance and creating the photometric humps. 
The inner edge of the hot disk had not yet reached the central white dwarf 
on the first day. That is a phenomenon well known as "UV delay" in
the dwarf nova outburst. This demonstrates that the photometric humps already 
observed at the very first day do not require any irradiation of the disk 
at all. 

Furthermore, we argue that the irradiating radiation from the
central source 
most likely did not penetrate deeply into the photosphere of the accretion 
disk as discussed in Sect. 3. In fact, higher Balmer lines 
H$\gamma$ and H$\delta$ and neutral helium lines, i.e., lines of the 
photospheric origin, were found in absorption even on July 24 
and on July 28  when He II at 4686 $\rm{\AA}$ and C III/N III complex 
were in emission (Baba et al. 2001) while H$\alpha$ and H$\beta$ 
lines exhibited highly variable absorption cores together with emission wings. 
This means that the heating source for continuous radiation must be  
deeply seated within the accretion disk rather than coming from
the outside 
(as in the case of irradiation). The continuum radiation is thus most 
likely due to the tidal dissipation, an argument favoring 
our tidal dissipation model against Kato's irradiation model 
for the early humps.

Let us now turn our attention to the problem of the radial range of the disk 
at which the tidal pattern is produced in Kato's model and in our model. 
Besides a localisation at the tidal cut-off radius, Kato (2002) considers
Ogilvie's (2002) vertical resonance as an explanation of the 
two-armed arch-like pattern observed in Doppler maps. The vertical 
resonance radius, $R_{\rm vertical}$, in the accretion disk is given 
by (see, e.g., Lubow 1981; Stehle and Spruit 1999)
\begin{equation}
R_{\rm vertical}/a= (1-\sqrt{1+\gamma}/2)^{2/3} (1+q)^{-1/3},
\nonumber
\end{equation} 
where $\gamma$ is the adiabatic exponent between pressure and density 
for the disk matter, $a$ is the binary separation, and $q=M_2/M_1$ is 
the ratio of secondary star to primary star mass. It is
interesting to note that this expression has the same mass-ratio
dependence as the 3:1 resonance for the eccentric tidal mode and the 2:1 
resonance for the Lindblad resonance only differing in the numerical 
factor. We estimate $R_{\rm vertical}/a= 0.323 (1+q)^{-1/3}$ 
for $\gamma=5/3$, and $0.441$ for $\gamma=1$,
while $R_{3:1}/a= 0.481 (1+q)^{-1/3}$ for the 3:1 eccentric resonance and 
$R_{2:1}/a= 0.630 (1+q)^{-1/3}$ for the 2:1 Lindblad resonance.  
Thus for a reasonable value of $\gamma$, the vertical resonance radius 
lies well inside the 3:1 resonance radius. 

If the two-armed arch-like pattern observed in Doppler maps of the 2001 
outburst of WZ ge were produced by the vertical resonance 
as suggested by Kato (2002) and if the disk in the 2001 outburst of WZ 
Sge were truncated at this radius, the disk would never reach the 3:1 
resonance radius and the common superhump phenomenon would not be expected. 
This is another difficulty of Kato's model. 

A further difference between  Kato's and our model lies in the 
expected absolute value of the velocity for the two-armed arch-like pattern in the Doppler map. The peak separation of double emission lines 
in quiescence of WZ Sge is known to be around 700 km/s 
(e.g., Mason et al. 2001). 
From  phase-folded spectra or Doppler maps of He II and
H$\alpha$ emission lines shown in Fig. 3 of Baba et al. (2002)
we find that 
the velocity of the peak-intensity features during the early-hump era of the 
2001 outburst of WZ Sge is around 500 km/s. If this velocity 
represents the Keplerian circular velocity, the outer edge of the disk 
expanded by a factor of about two during the outburst. Since the orbital 
velocity of the secondary star of WZ Sge was found to be around 500 km/s  
(Steeghs et al. 2001b) as well, the observed velocity of around 500 km/s 
for the two-armed features in Doppler maps means that they are produced 
very near to the Roche lobe of the primary star, much favoring the
2:1 resonance interpretation over that of the vertical resonance. 

We note that the observational accuracy does not allow to discuss a
difference between the radii of the 2:1 resonance and the ordinary tidal
cut-off, since for very small mass ratios these two radii can lie very close
together.

\end{appendix}

\end{document}